\documentclass[prb,preprint]{revtex4}

\usepackage{graphicx}

\begin{document}

\title{An undergraduate study of harmonic and parametric motion of
  a simple spring-mass system from motion waveforms}
\author{I. Boscolo}
\author{F. Castelli}
\author{M. Stellato}
\email{marco.stellato@unimi.it}
\affiliation{Dipartimento di Fisica, Universit\`{a} degli Studi di Milano}
\affiliation{Istituto Nazionale di Fisica Nucleare---Sezione di Milano,
via Celoria 16, 20133, Italy}

\date{\today}

\begin{abstract}

The spring-mass system studied in undergraduate physics laboratories may show complex dynamics due to the simultaneous action of gravitational, elastic, and torsional forces, in addition to air friction.  In this paper, we describe a laboratory exercise that caters to beginning students while giving those with more background an opportunity to explore more complex aspects of the motion. If students are not given predefined apparatus but are allowed to design the experiment setup, they may also learn something about physics thinking and experimental procedure.  Using results thus produced, we describe a variety of spring-mass oscillation patterns, discussing the physics of the significant deviations from simple harmonic motion. The parametric oscillation behavior we have observed is reported and investigated.  This study is based on analysis of motion waveforms.

\end{abstract}

\maketitle


\section{Introduction}

A mass hanging from a spring is a common, easy-to-perform experiment often used to introduce first-year physics students to simply harmonic motion.\cite{ib-lajpe}  Although the apparatus is simple and inexpensive, under certain conditions it turns out that the motion is not simple at all.\cite{olsson,cayton,christ,geba,gallo,arms,cush}
Gravitational (pendulum), elastic (spring), and torsional forces
\cite{olsson,cayton,geba,cush} generate numerous, complex phenomena that result in surprising motions in the spring-mass system.
These phenomena include multimode operation (many oscillating modes can be simultaneously active),\cite{cush} parametric instability (oscillation instabilities when the system is driven at certain frequencies), and energy transfer between the spring-bouncing mode and the pendulum-swinging mode.
A real spring-mass system behaves as a simple harmonic oscillator only under specific conditions:
(i) the spring's mass must be negligible compared to the
attached mass; (ii) the frequency of elastic oscillation must not
resonate with the frequency of pendular swinging; and
(iii) initial spring stretch must be strictly vertical.

A spring-mass system assembled with different values for mass and for spring constant can occasionally fall into a configuration where its elastic oscillation
frequency $\omega_k$ and its pendulum-oscillation frequency
$\omega_p$ have a ratio of nearly two.  In such cases, the motion is unstable
and the mass passes from vertical to horizontal oscillation in apparently random fashion, showing parametric rather than harmonic oscillations.
(A parametric oscillator is a harmonic oscillator that has a parameter
oscillating in time.)
In the spring-mass system, the gravitational force acting along the axis of the spring varies periodically with the pendular motion of the mass. This periodic action results in an exchange of energy and much more complex motion. The vertical oscillation amplitude decreases while the pendular amplitude increases, and vice-versa.

We noticed that an interaction between vertical and pendular oscillations sometimes occurred in our laboratory when students are free to choose the springs and masses themselves.  The physical aspects of the configurations that are linked to parametric behavior required further research to understand.  Thus, we performed a study of motion waveforms for a variety of frequency ratios that spanned the resonance value
$\omega_k/\omega_p =2$. The frequency ratio is selected at will
according to the equation
\begin{equation}
\label{ratio}
\frac{\omega_k}{\omega_p} = \frac{\sqrt{k/m}}{\sqrt{g/\ell}} \,,
\end{equation}
where $k$ is the spring constant, $m$ is the appended mass, $g$ is the
gravitational field strength, and $\ell = \ell_0 + mg/k$ is the equilibrium length of the vertical oscillating spring,
$\ell_0$ being the natural (unstretched) spring length.

In order to perform this experiment, students must deal with
the gap between theory and practice, examine complex, multi-effect
motion, and master experimental techniques.  Moreover, they have
to treat data statistically and purge the experiment of
many ``nuisances.''  In other words, the students are forced to
develop the ability to manage unexpected experimental
observations. The complexity of a phenomenon that is not fully understood requires a strategy
for singling out the motion components, relaxing their interrelations
so as to treat each component separately.  Afterwards, these components can be combined to understand the complete motion.  In short, this experiment turns out to be a useful, guided-research activity for students.

The theoretical work starts with the idealized equation of motion for
the deviation $z(t)$ from the (vertical) equilibrium position for a mass on a spring
\begin{equation}
\label{eq-freemotion}
\frac{d^2 z(t)}{dt^2} + \frac{C}{m}  \frac{d z(t)}{dt} +
\frac{k}{m}  z(t)  =  0  ,
\end{equation}
where $C$ is a damping coefficient.
The experimental goal is to test the solution
\begin{equation}
\label{z(t)-freemotion}
z(t) \,  =  \, z_0 \, e^{-\gamma\, t} \, \cos(\omega\, t),
\end{equation}
where $z_0$ is the initial oscillation amplitude,
$\gamma=C/2m$ is the damping constant, and
\begin{equation}
\label{eq-II}
 \omega^2 = \frac{k}{m} - \left(\frac{C}{2 m}\right)^2
= \omega_k^2  -  \gamma^2  .
\end{equation}
In these equations, the mass of the spring is assumed to be zero,
which is not actually the case.
The damping force is assumed to be proportional to velocity, although air friction against a moving object depends on its shape and is not (generally) linear with velocity.
The investigation proceeds through the following steps: (a) test of
Hooke's law $F = -k\, \Delta z$, (b) test of $\omega^2 = k
/m$, (c) measurement of the damping time $\tau = 1/\gamma$,
and (d) test of the sinusoidal solution Eq.~(\ref{z(t)-freemotion}).

In the following sections, we describe the procedure followed in laboratory work with students and point out its critical steps. In the final section, we present the experimental research work we carried out in preparation for teaching the course.  This study is meant to provide a deeper grounding for those conducting a widely used experiment; it is also a possible topic for open-ended investigation or small, individual research projects.  Complete investigation of the spring-mass system also requires studying forced oscillations,\cite{ib-lajpe} which is to be the subject of further research.


\section{The setup}

\begin{figure}[bt]
\centering \includegraphics[width=6cm]{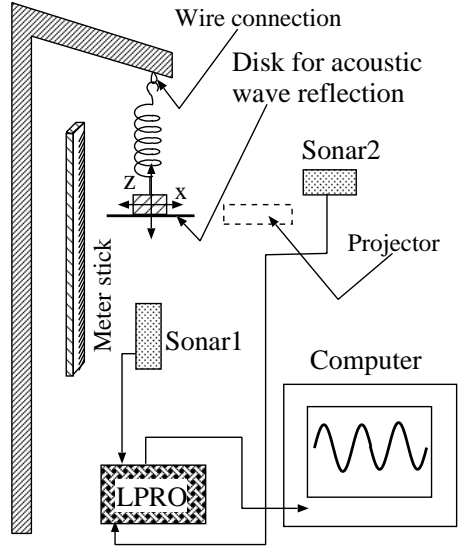}
\caption{\label{MarcoStellatoFig01}
Experiment layout. A meterstick and a light projector are
  used for measuring spring extension, while motion detectors (sonars) detect the
  motion. The digital data-acquisition tool LPRO (Vernier
  Logger Pro) acquires and sends the coordinates $z(t)$ and $x(t)$ to the
  computer. The top arm is bent to avoid reflecting the
  sonar signal.}
\end{figure}
The apparatus was kept as simple as possible: just a spring and a
weight oscillating vertically as shown in Fig.~\ref{MarcoStellatoFig01}.
A short piece of wire acts as a pivot. The $z$- and $x$-coordinates of the hanging mass,
defined in relation to its equilibrium position, are tracked by the bottom (Sonar1) and lateral (Sonar2) ultrasonic motion detectors, respectively.
The system is surrounded by plastic foam to absorb the direct sonar waves not reflected by the mass-bob.  The latter is a pile of 20 g metal disks on a support, to which plasticine may be added to obtain the desired precise weight.  The support is terminated by a diskette of 70-mm diameter to reflect the sound waves emitted by the bottom sonar as much as possible.  This technique is necessary because the mass has wide lateral oscillations and wobbles around its pivot.  Another spurious motion can be observed, namely: transverse spring vibration, possibly due to the mass wobbling.  The value of the diskette diameter is a compromise between a larger size for ideal sonar detection and a smaller size in order to minimize both friction and spurious motion.
These haphazard motions, superimposed on the bouncing and pendular
oscillations, cause observed motion waveforms to be irregular.
Distorted waveforms prevent us from attempting a simple analysis of the experiment and of its physics content.  To minimize spurious motion, it is essential to start with smooth, precise initial mass displacement and to use small oscillation amplitudes.

\begin{figure}[!bt]
\centering\includegraphics[width=8cm]{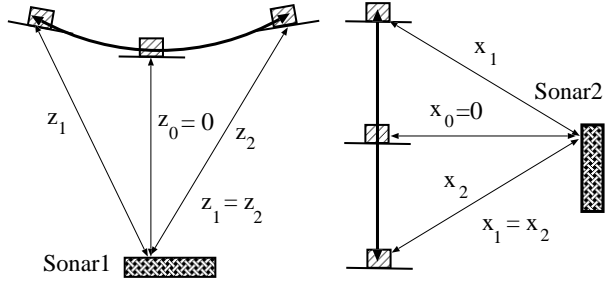}
\caption{\label{MarcoStellatoFig02} Scheme of the sound-wave paths
  between sonars and appended masses going through transverse
  motions. Sonars measure the to-and-fro component of movement.
  Transverse motion is seen as a difference in length between the two
  triangle sides connected to a sonar. }
\end{figure}
The two sonars ought to measure vertical and transverse oscillation exclusively, but this is not always the case.  Sonar1, which is used to detect vertical oscillation, partially detects transverse, pendular oscillation as well, as sketched in Fig.~\ref{MarcoStellatoFig02}. Similarly,
Sonar2, meant to detect the transverse motion, also detects
vertical spring oscillation.  This leads to waveforms in which an
oscillation with smaller amplitude is superimposed on the main
oscillation.  The resulting distortion of the oscillation under study can be observed by restricting the oscillations to pure motions. For pendulum motion, the spring was replaced with a wire; for vertical motion, a thin metal stick was inserted along the spring axis. The observation of pure motions is useful as a basis to interpret the results of combined oscillations described in the following sections.

It is worth mentioning that vertical motion detection by Sonar1 is not affected by the rotation of the pendular oscillation plane, but this is not true with Sonar2. When the oscillation plane rotates by large
angles, Sonar2 waveforms become almost useless.  Good waveforms from
Sonar1 are enough to study the motion, with the occasional help of some information from Sonar2.

A meterstick and a stopwatch are used to measure elongations and oscillation periods, respectively. Typical values of the spring parameters are $m_{\rm spring}\simeq 4.8$\,g,
$k\simeq 8$\,N/m, and rest length $18$\,cm. The effective mass values for the bob---equal to the hanging mass plus a third of the spring's mass\cite{ib-lajpe,christ,cush,halliday} (discussed more fully below)---range
from 40--80\,g.

In class, students assemble the system using the bottom Sonar1 only.  This paper consists mainly of research performed after the students had completed the lab, and was carried out to investigate their unexpected results.


\section{Test of Hooke's law and $\omega^2 = k /m$}

In our simple harmonic motion experiment, students determine the spring constant in the traditional way. A test of Hooke's law $F = -k\, \Delta z$ is performed by adding different masses to the spring and measuring the relative static elongations using a meterstick.  The spring constant is obtained with a precision of a few percent.
The test of $\omega^2 = k /m$ is performed using masses ranging from a maximum value, limited by spring damage, to a minimum value, determined by excessive vibrations. Some oscillations turn out to be very irregular, particularly when smaller masses are used. Hence, students
are guided towards using a larger mass to obtain more regular oscillations. The oscillation period is measured using a stopwatch.  The initial objective was to verify $\omega^2 = k /m$, rewritten in terms of the period $T$ as
\begin{equation}
\label{m-T-eq}
 m  =  \frac{k}{4 \pi^2} T^2.
 \label{emme}
\end{equation}

\begin{figure}[!bt]
\centering\includegraphics[width=8cm]{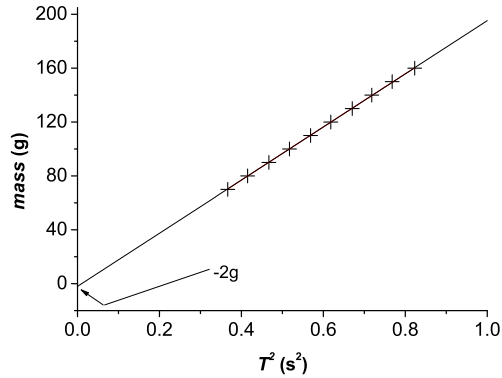}
\caption{\label{MarcoStellatoFig03} Graph of Eq.~(\ref{eq-mass}), showing the
  negative intercept at the $y$ axis.}
\end{figure}

A graph of the added mass versus $T^2$ shows the expected straight line
with a small negative intercept on the $y$-axis, as shown in Fig.~\ref{MarcoStellatoFig03}.
This test shows the effect of the non-negligible mass of the spring.
A class discussion of the situation \cite{ib-lajpe} leads to
the conclusion that the spring's mass cannot be neglected, and that the added mass should be
replaced by an effective mass that can be written $m_e = m +m_s$, where $m_s$ is the effective contribution of the spring's mass.
Hence, Eq.~(\ref{emme}) must be rewritten as
\begin{equation}
\label{eq-mass}
m  =  \frac{k}{4\pi^2} T^2 - m_s .
\end{equation}
The $y$-intercept from the data then gives the expected value of one-third of the total spring mass
($ m_s = m_\mathrm{spring}/3$), as reported in the literature.\cite{christ,halliday}  For the remainder of
this paper, the pendulum mass will actually refer to the effective mass as defined here unless stated otherwise.

While students notice that lighter masses cause more complex oscillations, they also realize that the system does not show the expected simple harmonic oscillation and ask assistants for help. The new phenomenon of parametric oscillations comes to the fore for the first time in this context, and the complexity of the movement leads to further measurements with multiple motion detectors connected to a computer.


\section{Decay time and frequency measurements using motion waveforms}
\label{stau}

After completing the previous tests, students start taking
measurements for the decay time $\tau = 1/\gamma$ in Eq.~(\ref{z(t)-freemotion}).  The time interval between the beginning of the oscillation and the moment
when the amplitude is reduced to $1/e$ of its initial value is
measured using a stopwatch.  Students use larger masses to get reasonably stable oscillations. They observe that $\tau$ depends on initial oscillation amplitude and on the value of the added mass (as expected from the definition $\tau=2m/C$).

At this point, the study of the motion using waveforms is introduced,
with the aim of measuring the two parameters $\tau$ and $\omega$ directly from the computer screen to check their previously obtained results. In theory, the value of decay time is constant for a given mass; nevertheless, we measure it for different sections of the same decay curve obtaining
different values.  Indeed, the decay does not appear to be exponential towards the end of the curve, instead showing decay times that increase with time.  The very slow decay at the tail indicates an effective reduction of the damping coefficient; that is, a reduction of the effect of air resistance when the mass oscillates more slowly. As a result, the mathematical form of Eq.~(\ref{z(t)-freemotion}), based on a constant decay value, does not reproduce actual, observed motion.  Initial fast decay indicates the possibility of energy transfer from the vertical motion to other motions.  For the sake of thoroughness, we add that the assumed linear dependence of the damping time $\tau$ on the hanging mass ($\tau=2m/C$) was tested, with better than 80~percent agreement. \cite{ib-lajpe}

After recording the variation in decay time along the waveform curves on the screen, students use the same graphs to measure oscillation frequency.
They determine the time interval relative to a set of oscillations on the computer screen, then use a fast Fourier transform (FFT) tool, part of their data acquisition software, which provides the frequency spectrum.  In this
context, students are introduced to the concept and use of the FFT tool.
After this measurement, students take a look at the waveforms obtained
with different masses.  The seemingly harmonic behavior with heavy masses,
the very complex motion with elastic vertical and transverse pendular
oscillation interchange, and the strong variation of waveform decay at the beginning of the motion make it clear that a new and more complete
model for the system is needed to account for the many discrepancies between
the simple motion predicted by the theory of harmonic oscillation and actual experimental observations at variance with expected results.

Separately, a few highly-motivated students then further studied the discrepancy between theory and experiment by fitting the
experimental waveform to the function $f(t)= a(0) \exp(-t/\tau) \sin(\omega_0 \, t + \phi) + b$.
The values of $a(0)$, $\tau$, and $\omega_0$ (initial amplitude, decay time, and oscillation frequency, respectively) are extracted from the waveform as explained above.
Fitting the whole waveform turned out to be impossible; reasonably good fits were obtained only within sections of the decaying waveform.  The farther the waveform section the longer the relative $\tau$.

As an example, the fit of a mathematical model to a waveform using a 70 g mass gives the following results, which are in accord with the previous conclusion:
(1) $\tau \sim 12$\,s for the first 10\,s section, (2) $\tau \sim 17$\,s for the first 20\,s (a longer section averages two different measured decay times),
(3) $\tau \sim 20$\,s for the curve section $\Delta t = 10$--30\,s,
(4) $\tau \sim 30$\,s for the curve section $\Delta t = 20$--60\,s, and
(5) a much longer result at the curve tail.
We checked the perfect correlation between the $\tau$ enhancement at a curve section and the relative frequency enhancement at that curve section, as expected theoretically from Eq.~(\ref{eq-II}).


\section{Measurements on harmonic and parametric motions}

\begin{figure}[!ht]
\centering \mbox
{\includegraphics[width=8cm]{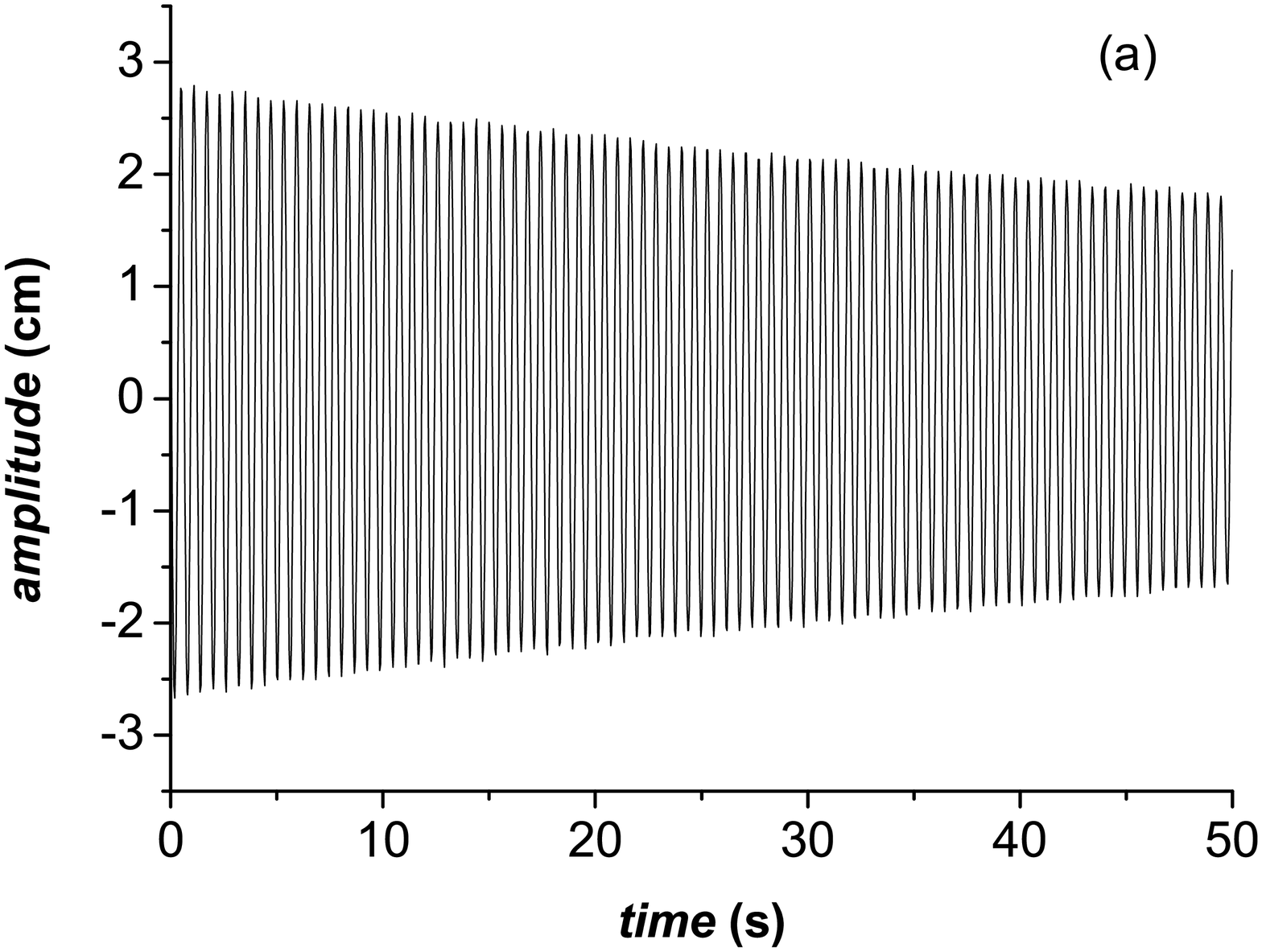}}
{\includegraphics[width=8cm]{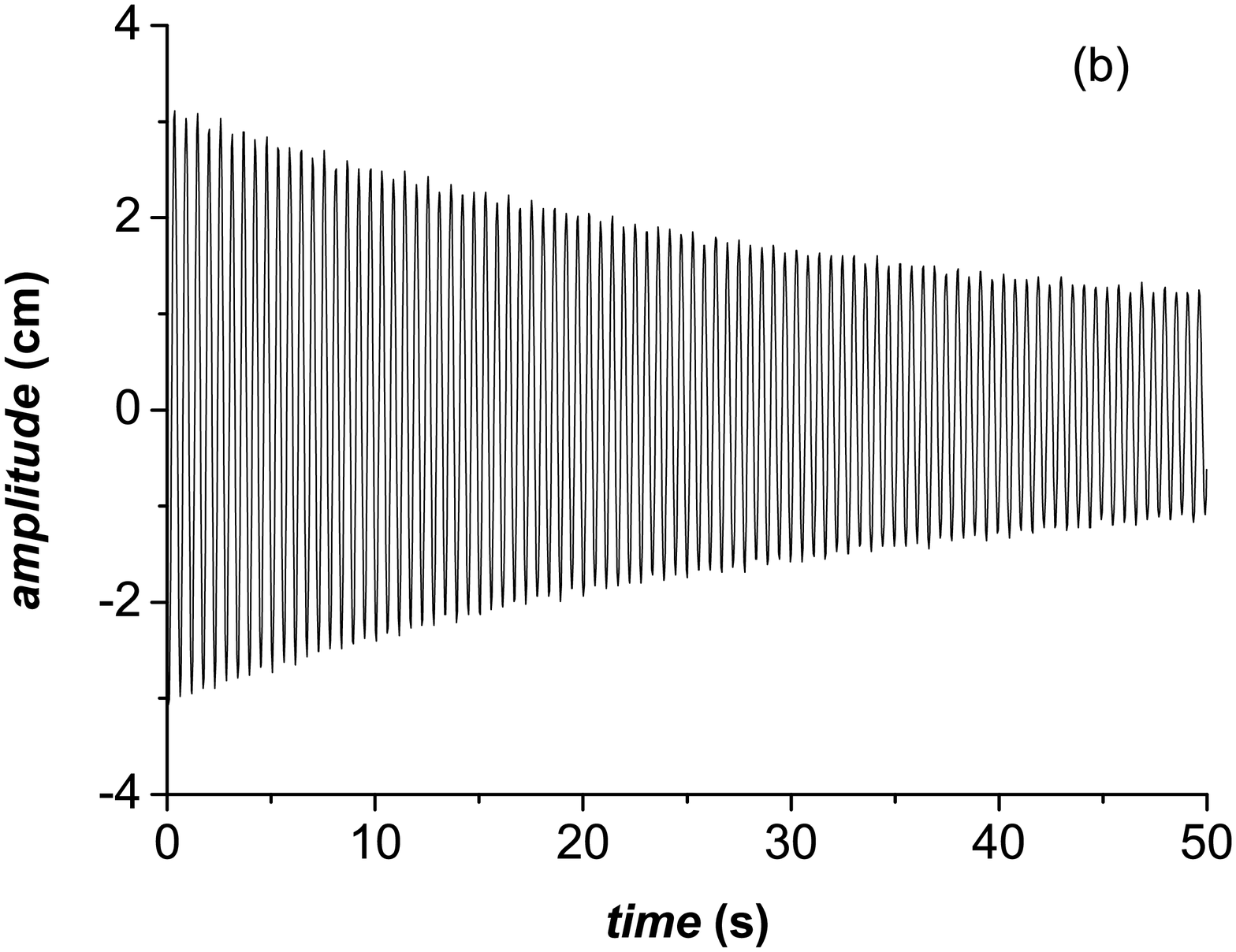}}
{\includegraphics[width=8cm]{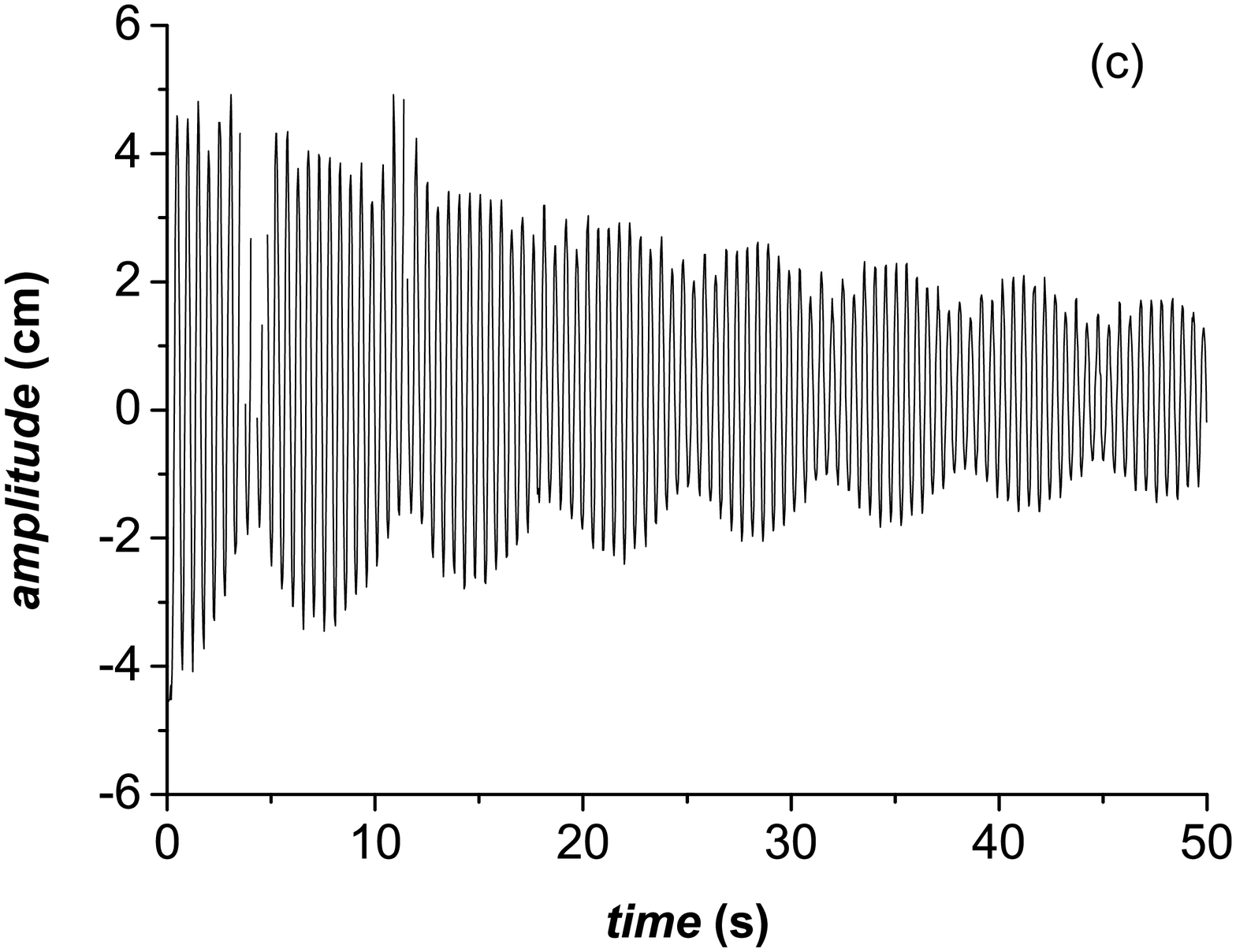}}
{\includegraphics[width=8cm]{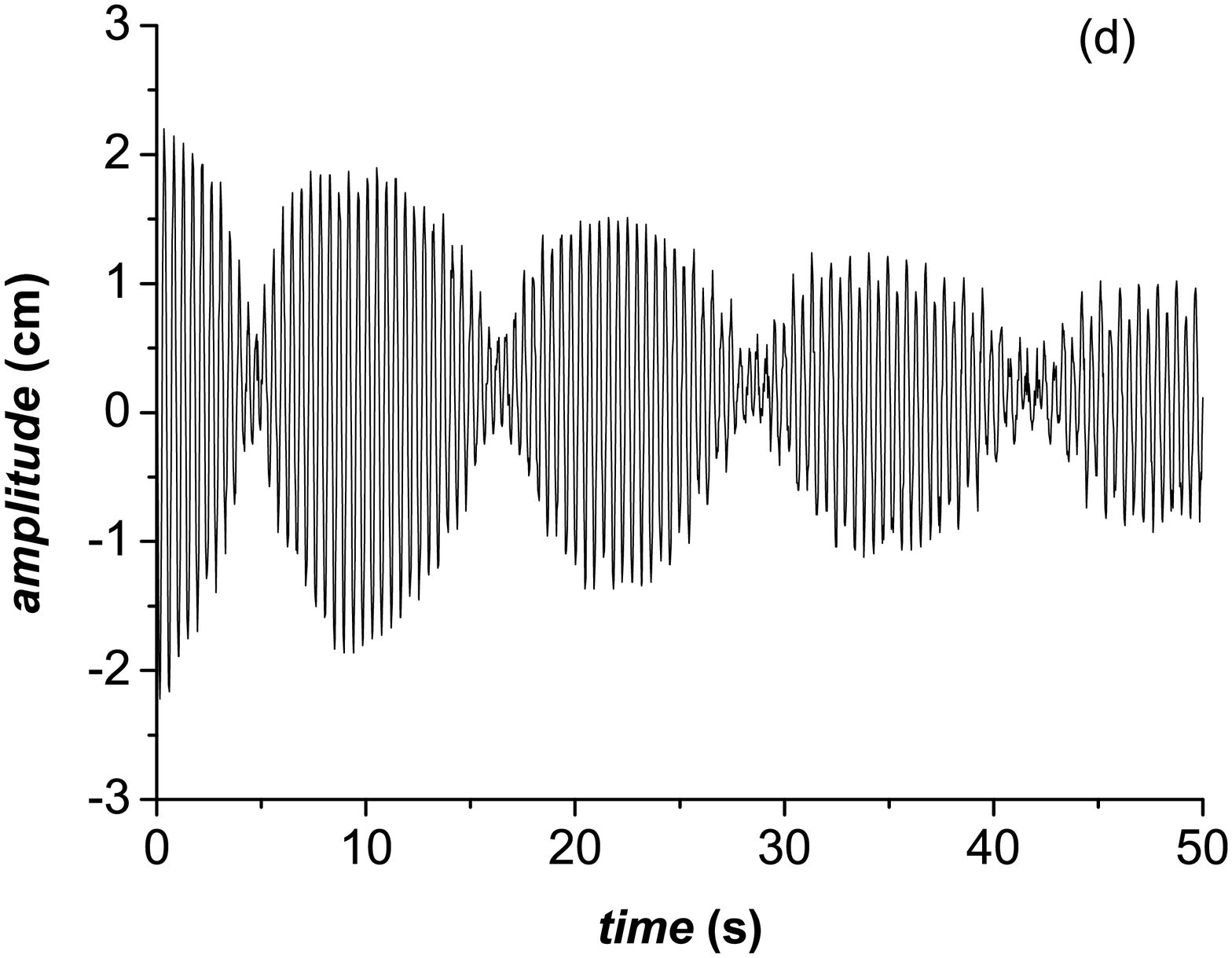}}
\caption{\label{MarcoStellatoFig04} Waveforms (a), (b), (c), and (d)
  corresponding to 70~g, 60~g, 50~g, and 40~g masses, respectively. The system passes from out-of-resonance, with the 70~g mass and $\omega_k/\omega_p\simeq
  1.67$, to near-resonance, with the 40~g mass and $\omega_k/\omega_p\simeq 2.04$.  The waveforms evolve from a classical, damped sine wave towards complete modulation.  The comb-like crests along the signal envelope,
  evident in frame (d), are generated by the particular way Sonar1 detects the transverse motion component.}
\end{figure}

The remainder of this paper describes details of faculty members' research
on motion coupling within the spring-mass system.
This analysis is beyond the level of first-year undergraduate teaching.  Typical waveforms obtained with a set of 70 g, 60 g, 50 g, and 40 g masses are shown in Fig.~\ref{MarcoStellatoFig04}.

Waveforms from heavier to lighter masses show a transition from an
exponential decaying sinewave to partially modulated decaying sinewaves and,
finally, to a multi-lobed, completely modulated sinewave.  The same developments are observed when looking directly at the motion, with composite vertical-pendular
motion for masses of intermediate value and the addition of
frequent switches between vertical and pendular oscillations with the
lightest mass.  The intermediate case of a modulated-edge sinewave
indicates that oscillation exchange between the two modes is partial.
By counting the lobes,
we can measure the frequency of oscillation-mode exchange.
Another interesting observation is that the plane of pendular
oscillation is not stable.  It rotates by a few degrees in the case of
low partial mode exchange. Conversely, when mode
exchange is complete, the oscillation plane shifts by larger angles (up to 90
degrees) in an apparently random manner.

The study of waveforms leads us to conclude that
there are two different classes of motion: harmonic and
non-harmonic. The two types of motions refer to different
physical phenomena.
\begin{figure}[!bt]
\centering\includegraphics[width=6cm]{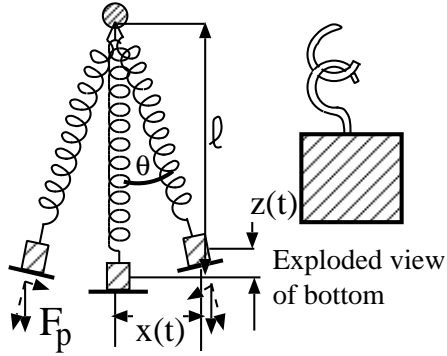}
\caption{\label{MarcoStellatoFig05} Sketch of the applied force showing coupling
  between spring-bouncing and pendulum-swinging motions. The mass' center of gravity is not along the spring axis.}
\end{figure}

Let us again look at the forces acting in the system, as depicted
in Fig.~\ref{MarcoStellatoFig05}.
We note that the component along the spring axis of the force $mg$ applied by the appended mass changes periodically during
pendular oscillation and, in turn, alters spring extension.
In particular, stretching-force oscillation causes the spring's equilibrium
length to oscillate around its initial value $\ell$:
\begin{equation}
\ell(t) = \ell +\delta \ell(t).
\end{equation}
Therefore, pendular and spring-oscillation modes are coupled.
Spring extension due to the weight makes a complete oscillation in half the time of one complete
pendulum oscillation at resonance, that is when the ratio between
the two oscillation periods $T_k$ and $T_p$ is 1/2.

The coupling between pendular oscillation and spring oscillation induces
parametric instability between the two modes \cite{olsson,cayton,lai,holz} when the resonance condition is met. The energy exchange between the modes, tested by the waveform envelope modulation, is observed up to a frequency ratio [Eq.~(\ref{ratio})] of about 1.8.  The farther from the resonance condition, the lower the energy exchange.  The observed mode exchange within an interval of frequency ratio reproduces the common physical fact that the resonance curves have
Gaussian-like form with a certain width.  We can say that our resonance
has a relatively large width.
Moreover, out of resonance, our measurements show that the greater the amplitude of vertical
oscillation, the more extensive mode exchange becomes, yielding a wider
resonance curve.
The resonance curve was also observed to enlarge as motion disorder increases, i.e. when
spurious motion becomes significant compared with the two motions of spring oscillation and pendulum oscillation.
Spurious motion is always present when oscillations are large. Both spurious motion and large oscillation lead to an increase of the mode coupling.

\begin{figure}[!ht]
\centering\includegraphics[width=8cm]{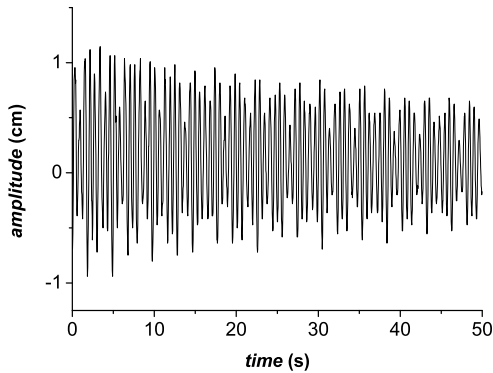}
\caption{\label{MarcoStellatoFig06} Waveform of vertical
  oscillation obtained after a lateral initial displacement.
  The irregular shape of the waveform depends on the simultaneous
  presence of vertical and transverse oscillations, as seen by
  Sonar1. A 70~g mass was used in this test.}
\end{figure}
Motion due to oscillatory mode exchange (parametric motion)
is initiated by either vertical
or lateral shift of the appended mass, i.e. with
either elastic or pendular oscillation excitation.
Vertical motion represented by a waveform like that of Fig.~\ref{MarcoStellatoFig06} is generated by
starting pendular oscillation with lateral displacement of
the appended mass.  This parametric motion suggests that the coupling between pendular and
elastic oscillations is nonlinear.  In the literature, equations
describing nonlinear coupled pendulum and spring oscillations are
given as\cite{olsson}

\begin{eqnarray}
\label{eq1-parametric}
\frac{d^2 x(t)}{dt^2} + \omega_p^2 \, x(t) &=& c \, x(t)\, z(t), \\
\label{eq2-parametric}
\frac{d^2 z(t)}{dt^2} + \omega_k^2 \, z(t) &=& c \, \frac{x^2(t)}{2} ,
\end{eqnarray}
where the coupling constant $c$ is a function of the system
parameters.\cite{olsson}  Incidentally, damping is not considered
in these equations.  More refined motion equations that satisfactorily
describe the results of our experiment are presented in Ref.~3. These equations can be obtained
through Lagrangian formulation.  This system can be solved analytically only in one particular case
\cite{olsson} and numerically in all others.\cite{cayton}.
\begin{figure}[!ht]
\centering \mbox
{\includegraphics[width=8cm]{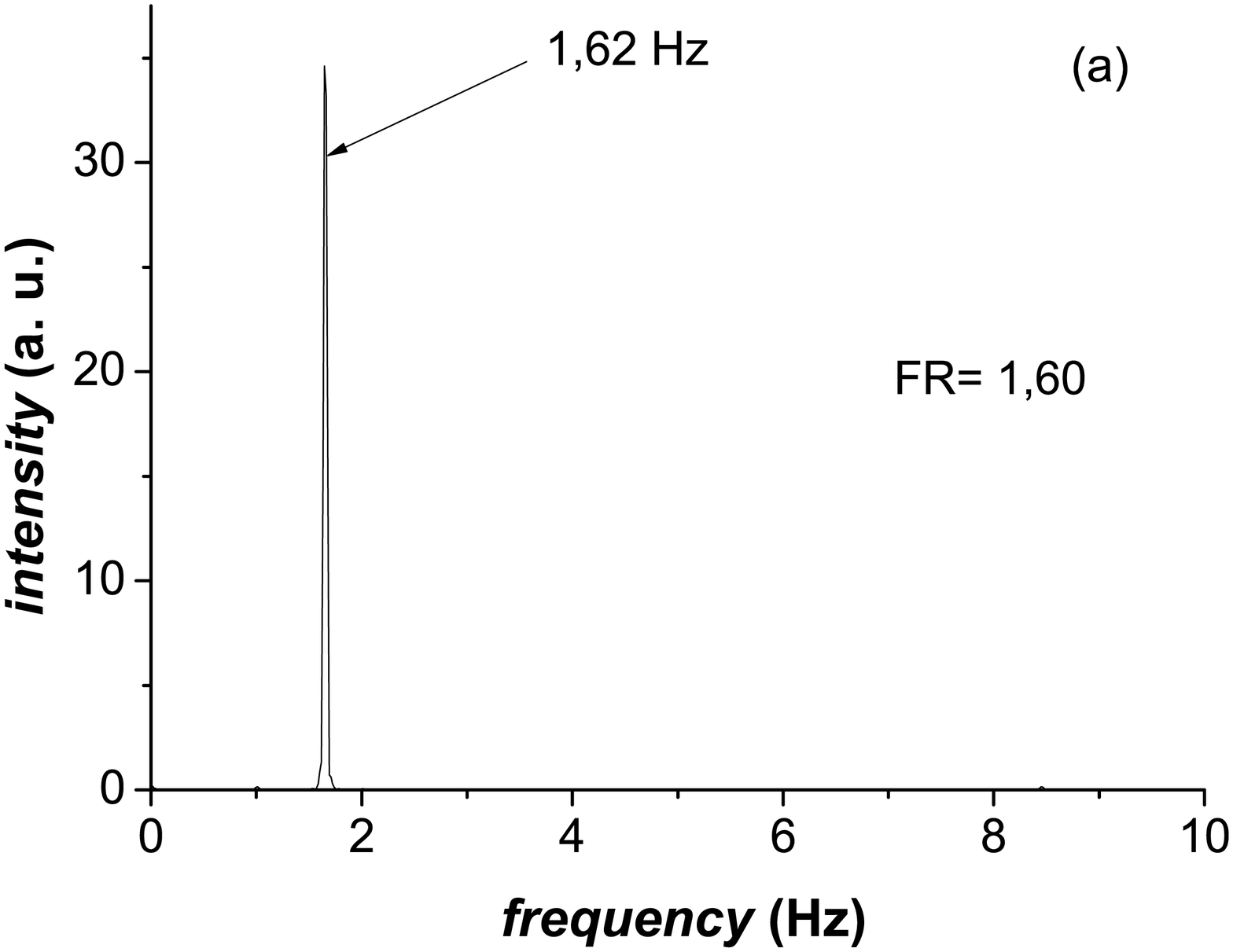}}
{\includegraphics[width=8cm]{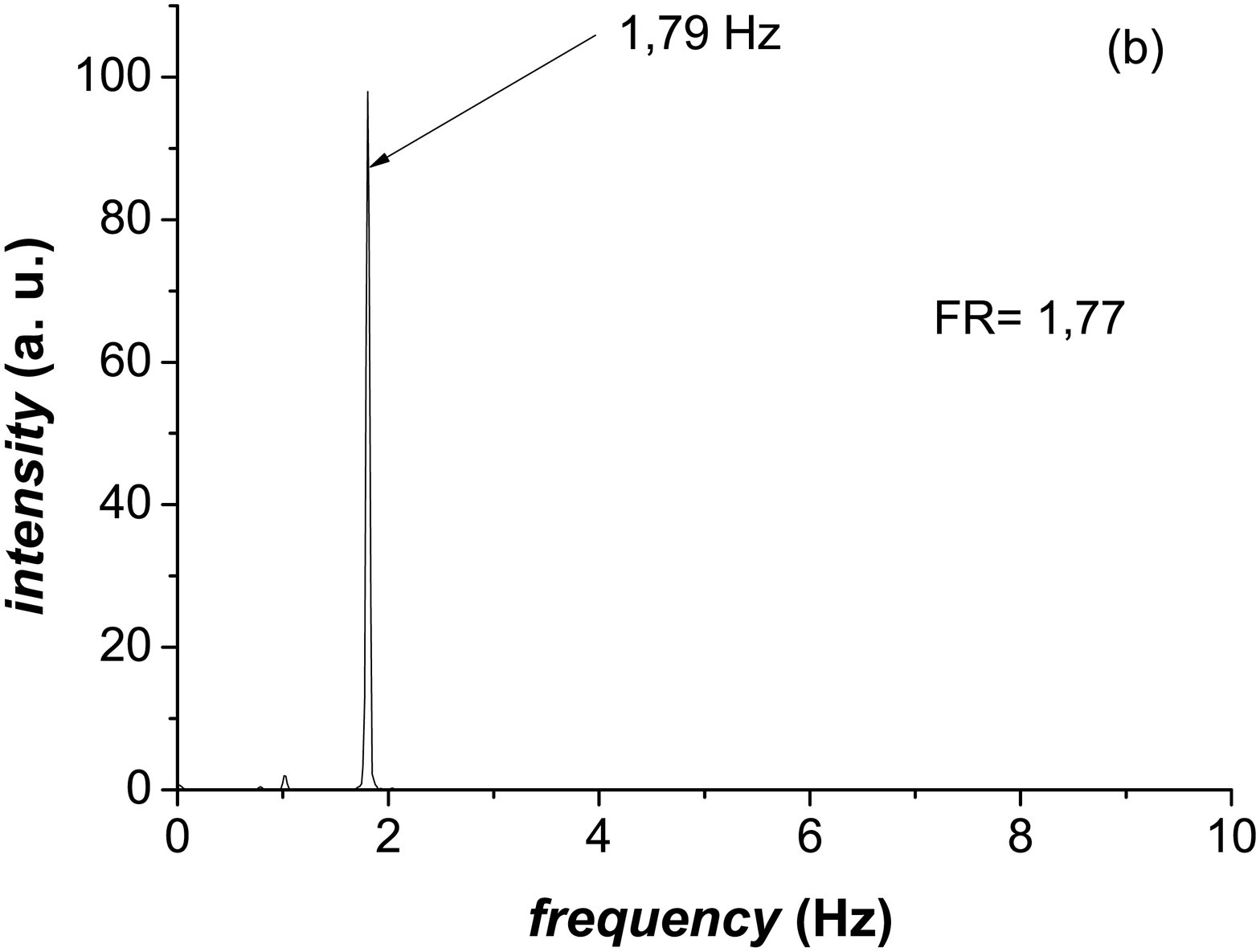}}
{\includegraphics[width=8cm]{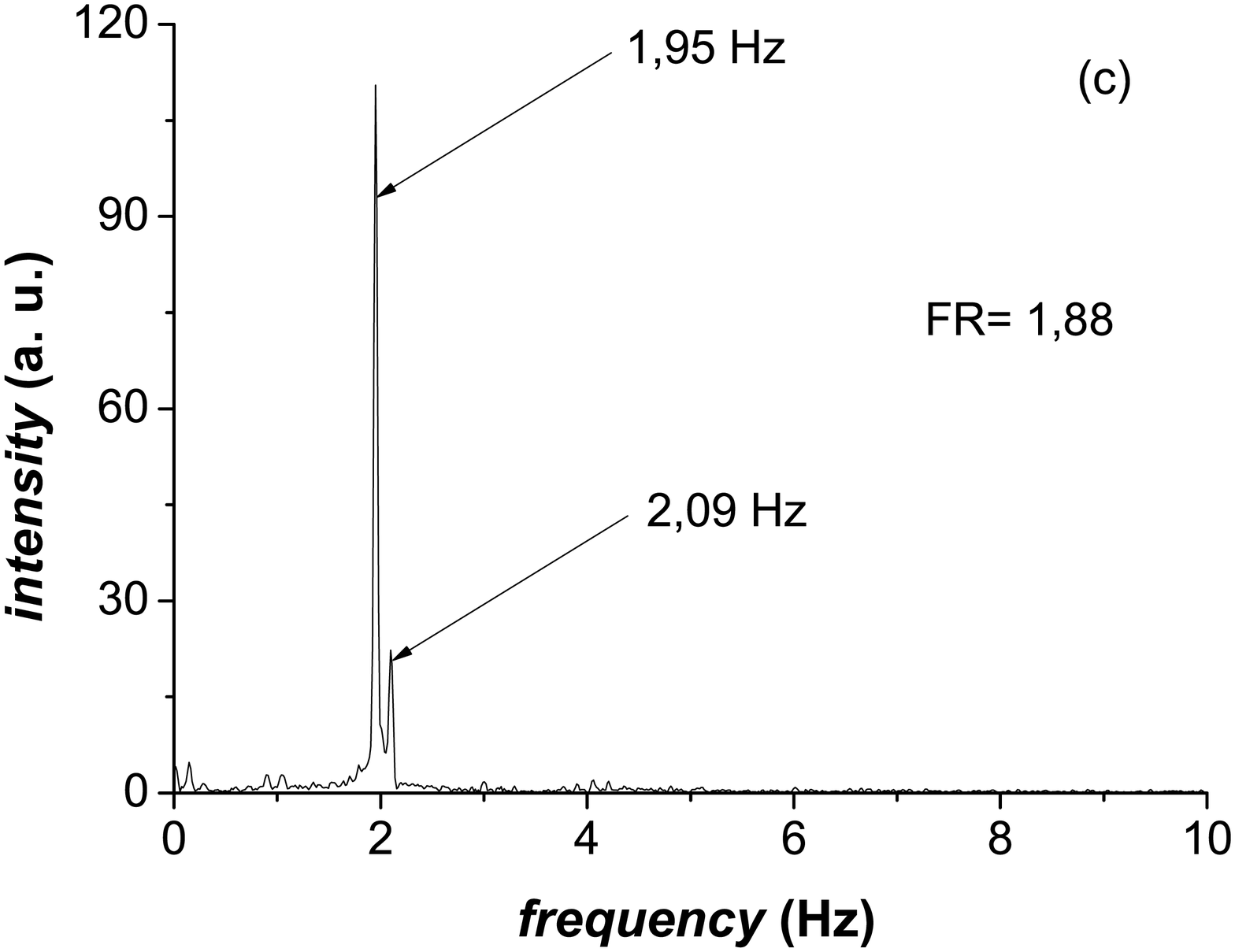}}
{\includegraphics[width=8cm]{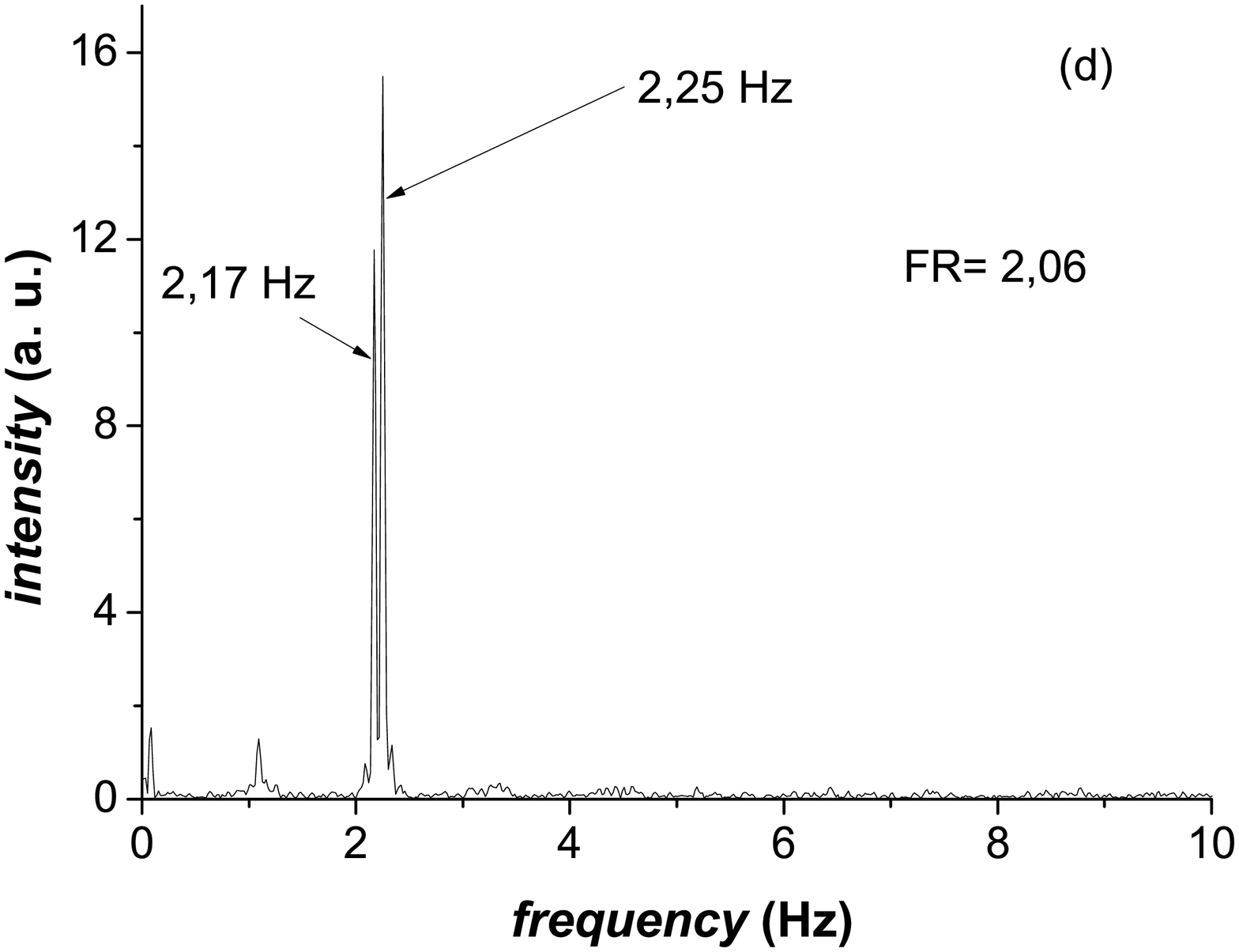}}
\caption{\label{MarcoStellatoFig07} Plots (a), (b), (c), and (d) show the
FFT signals derived from 70~g, 60~g, 50~g, and 40~g waveforms, respectively.
FR is the actual value of the frequency ratio. The frequencies of some pronounced spectrum lines are labeled.}
\end{figure}

The presence in the motion of different oscillation modes can be observed
directly in the waveform spectrum lines obtained with the FFT tool.
The spectra of the waveforms in Fig.~\ref{MarcoStellatoFig04}, shown in Fig.~\ref{MarcoStellatoFig07}, have the expected
frequency lines for the oscillations those waveforms represent.
The neater the waveform (i.e. the more ordered the motion), the neater the
FFT.  In each frame, the line corresponding to $\omega_k$ is largely dominant,
and the line corresponding to $\omega_p$ is present, albeit with limited intensity.
In disordered motion, the lines for spurious oscillation modes show up in the spectra.
The two (c) and (d) spectra of Fig.~\ref{MarcoStellatoFig07} make the $2 \omega_p$ harmonic stand out,
because the pendular oscillation has large amplitude and Sonar1 doubles the pendular frequency, as explained in Fig.~\ref{MarcoStellatoFig02}.

\subsection{Behavior with mixed vertical and transverse excitation}

Any excitation (initial mass shift with vertical and lateral components)
that corresponds to a frequency ratio within the resonance width can start
mixed vertical and pendular motion.
Indeed, the coupling term on the right-hand side of motion equations
(\ref{eq1-parametric}) and (\ref{eq2-parametric}) shows that
an initial $x(0) \neq 0$ causes the onset of energy exchange
between the two modes.  It is almost impossible to apply a
pure vertical or a pure lateral shift by hand.

\begin{figure}[!ht]
\centering
\includegraphics[width=8cm]{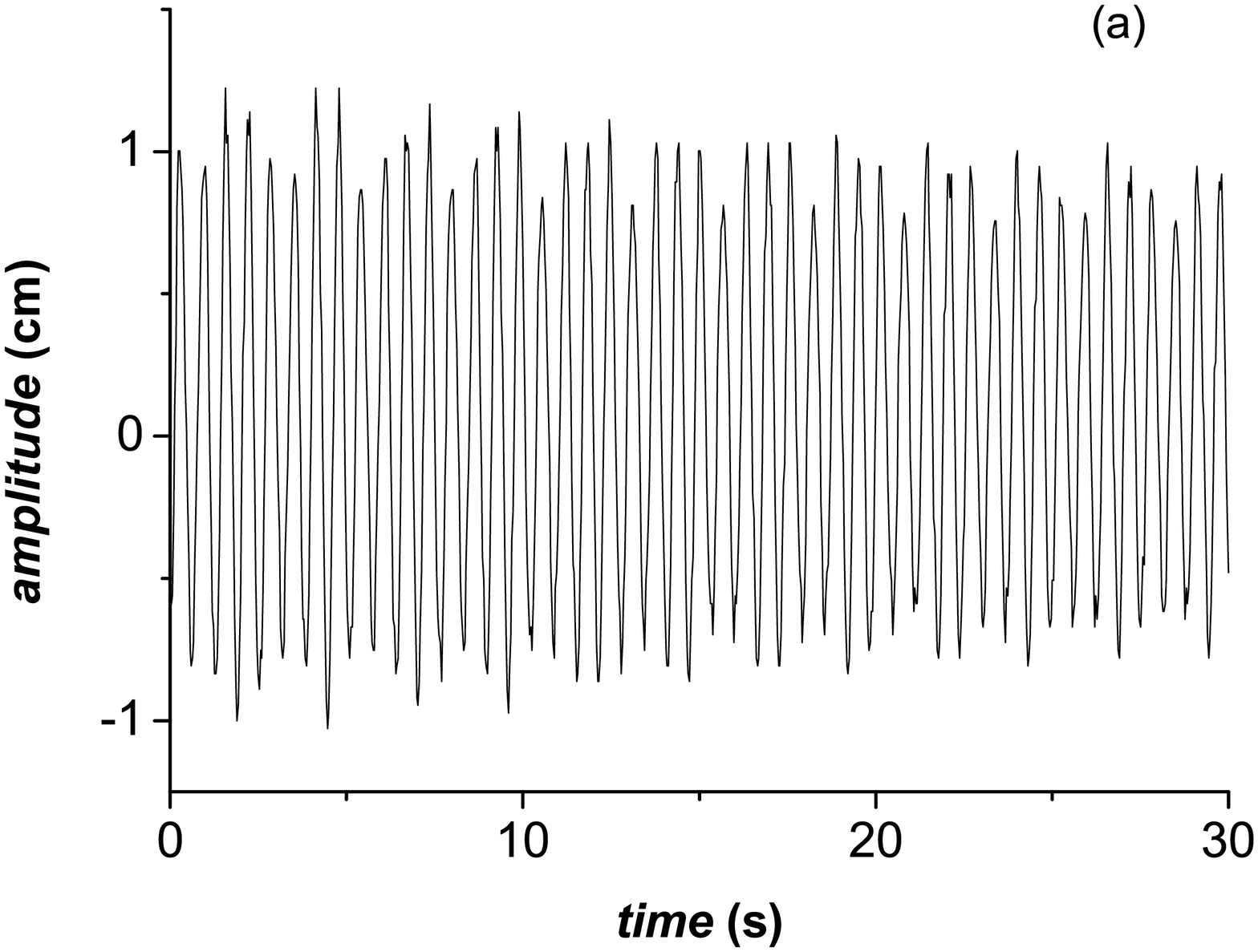}
\includegraphics[width=8cm]{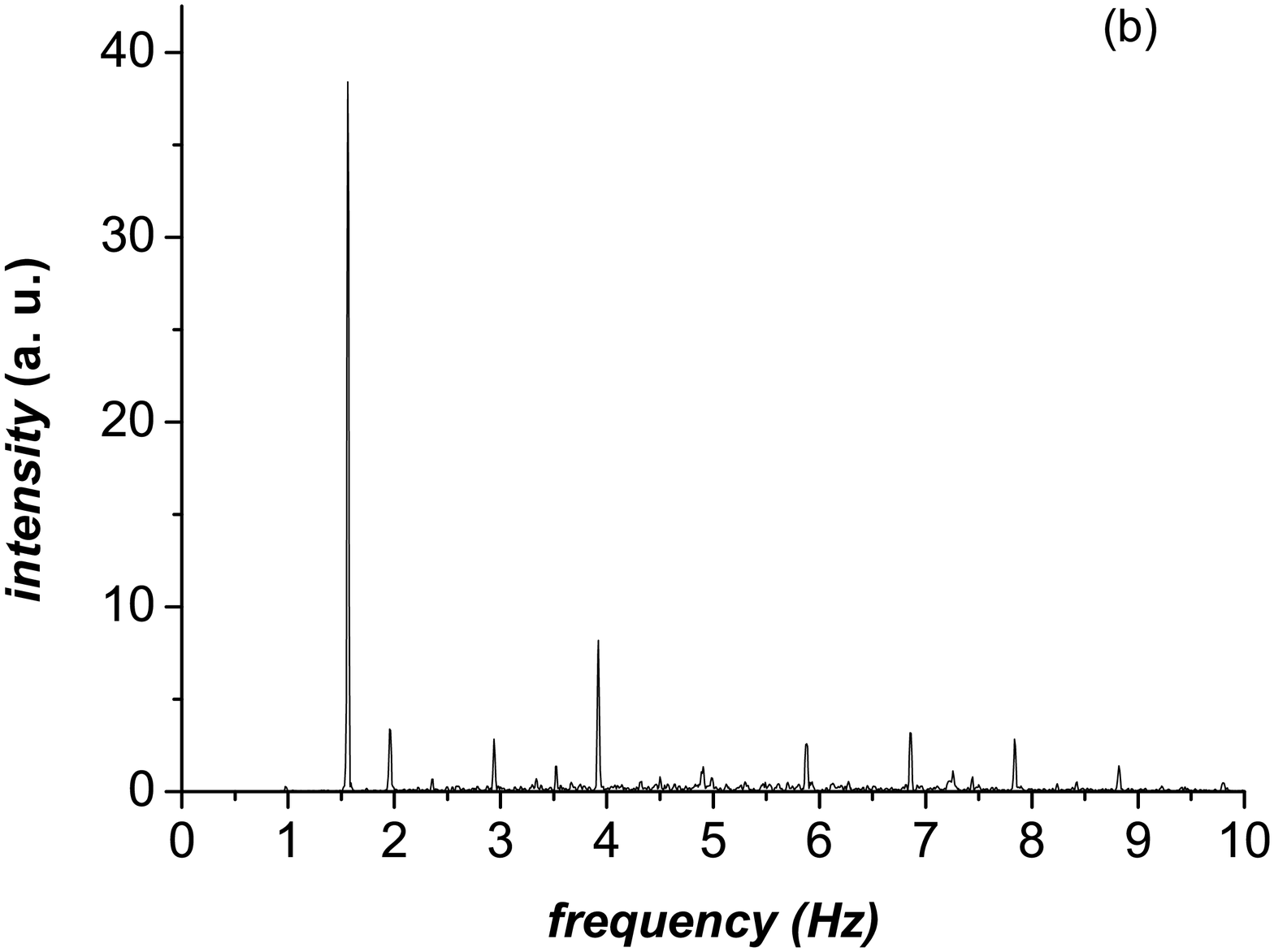}
\caption{\label{MarcoStellatoFig08}
Frame (a): Waveform obtained with an 80~g mass and small diagonal
  displacement. The energy exchange between the two modes is clearly visible.
  Frame (b): Frequency spectrum. Many lines are present,
  besides the spring frequency at 1.55 Hz and the pendular frequency at 0.95 Hz (with very small amplitude). Second harmonic frequencies are visible. Other clear lines, at higher frequencies, correspond to wobble motions.}
\end{figure}
\begin{figure}[!ht]
\centering\includegraphics[width=4cm]{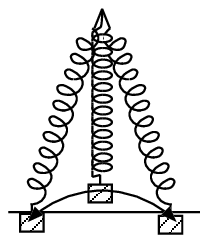}
\caption{\label{MarcoStellatoFig09} Motion figure normally assumed by masses
above 60~g. }
\end{figure}
The result obtained with mixed excitation is interesting: energy
bouncing between the two modes occurs even with
relatively heavy masses, i.e. with a frequency ratio that is out of resonance. The
waveform obtained with 80 g mass (1.63 frequency ratio) proves to be modulated
as shown in the first frame of Fig.~\ref{MarcoStellatoFig08}.
Its FFT spectrum has many sharp lines, as shown in the second frame of Fig.~\ref{MarcoStellatoFig08}.
A rather intriguing observation is that the system set itself in the peculiar
stable motion shown in Fig.~\ref{MarcoStellatoFig09} after a few oscillation cycles---a behavior cited also in Ref~2.
Results of mixed-excitation experiments enable us to
state that nonlinear coupling between the two modes occurs even far from resonance.

Other  observations worth reporting are:

\begin{itemize}

\item[(1)] At resonance, both vertical and lateral mass displacements lead
  to similar system instability.

\item[(2)] Near resonance conditions, both strong vertical and strong lateral
    mass displacements lead to substantially similar results in waveforms
  and spectra.  This is not the case with small displacements: a small
  vertical displacement excites only vertical spring oscillation (no exchange
  with pendular oscillation), whereas a small lateral displacement excites both modes.

\item[(3)] Mixed excitation produces complete energy transfer between
  the two modes even at the frequency ratio $\omega_k/ \omega_p=1.8$.
  The spectra are neater than in pure vertical excitation.

\end{itemize}


\subsection{Analogy to optics}

Parametric instability is common in nonlinear optics.  This phenomenon
is employed extensively to produce and amplify laser waves at diverse desired
frequencies.\cite{yariv,aegis}  A pump wave $\omega_1$ launched across a
nonlinear crystal generates, from noise, a signal wave $\omega_2$ along with a so-called idler wave $\omega_i$.  Among others, the main condition that
must be satisfied is
\begin{equation}
\label{tre-onde}
\omega_1  = \omega_2 + \omega_i \,  .
\end{equation}

\begin{figure}[!ht]
\centering\includegraphics[width=8cm]{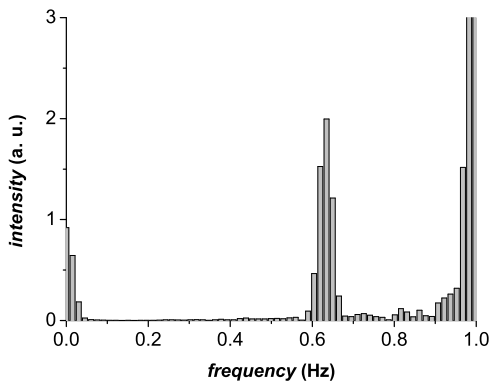}
\caption{\label{MarcoStellatoFig10} The idler frequency line at
  $ 0.62 = (1.63-1.01)$~Hz from the FFT graph of the waveform generated
  by vertically exciting a 70~g mass (see Fig.~\ref{MarcoStellatoFig07}(a)), recorded by Sonar2.}
\end{figure}

In our spring-mass system, the spectrum clearly shows the idler frequency,
\begin{equation}
\label{idler}
\omega_i = \omega_k-\omega_p  \, .
\end{equation}
For example, the spectrum of the waveform yielded by vertically exciting a
70~g mass shows the idler line distinctly, as
seen in Fig.~\ref{MarcoStellatoFig10}.  This figure reproduces a selected
section of the spectrum recorded by Sonar2, which, for this type of
excitation, detects low-amplitude components more cleanly.
%


\section{Conclusion}

The vertically oscillating spring-mass system used to study harmonic
oscillator physics in a first-year laboratory course has to be
pre-designed in order to reliably reproduce the motion described in
textbooks.  In a laboratory organized so that students can assemble
their own apparatus as they see fit (picking components from a pile),
the spring-mass system is not likely to turn out so that it reproduces
the idealized harmonic motion.  There is a good chance they will come
up with a system that shows complex dynamics.  Furthermore,
to test the spring-mass physics law $\omega^2=k/m$,
students have to apply different masses, thus often running into parametric
instability. Hence, studying the spring-mass system in a laboratory
where students pick the parts addresses both harmonic and parametric
behavior.  We believe the spring-mass experiment can be treated in
class as a harmonic oscillator and then developed towards the
parametric oscillator. This teaching sequence allows students to
approach the challenging, complex physics content of parametric behavior.

One of this paper's main contributions is its experimental investigation
of the parametric motion of the spring-mass system, as dealt with in
the first-year laboratory course for physicists.  The study was performed
by tracking the motion of the appended mass with an ultrasonic motion
detector and capturing the waveform on a computer screen.  Analysis of
the waveform clearly reveals the parametric instability and the energy
exchange between the bouncing spring and the swinging, pendular
oscillation, when the resonance conditions between the two oscillations
are met. Resonance width is measured by looking at waveforms
in a frequency interval around the resonance.

The rather complex---but manageable---experiment and the gap between the
ideal theoretical case and the actual experimental case make this
experiment a useful tool for stimulating an aspiring physicist's thinking.
We are working on a presentation of a more in-depth study of parametric
spring-mass physics.


\begin{acknowledgments}

We thank M. Giliberti for useful hints, and R. Lowenstein for supporting
this research and for critical reading of the manuscript.
We gratefully acknowledge technical assistance from D. Cipriani.

\end{acknowledgments}


\end{document}